# Revealing intrinsic vortex-core states in Fe-based superconductors through machine-learning-driven discovery

Yueming Guo[1*†], Hu Miao[2†], Qiang Zou[1,3†], Mingming Fu[1,4], Athena S. Sefat[2], Andrew R. Lupini[1], Sergei V. Kalinin[1*], Zheng Gai[1*]

[1]Center for Nanophase Materials Sciences, Oak Ridge National Laboratory, Oak Ridge, Tennessee 37830, USA

[2]Materials Science and Technology Division, Oak Ridge National Laboratory, Oak Ridge, Tennessee 37830, USA

[3]Department of Physics and Astronomy, West Virginia University, Morgantown, West Virginia 26506, USA

[4]Department of Physics, OSED, Fujian Provincial Key Laboratory of Semiconductor Materials and Applications, Xiamen University, Xiamen 361005, China

† These authors contributed equally.

* Corresponding authors: dr.yueming.guo@gmail.com, sergei2@utk.edu, gaiz@ornl.gov.

**Abstract**

Electronic states within superconducting vortices hold crucial information about paring mechanisms and topology. While scanning tunneling microscopy/spectroscopy(STM/S) can image the vortices, it is difficult to isolate the intrinsic electronic states from extrinsic effects like subsurface defects and disorders. We combine STM/S with unsupervised machine learning to develop a method for screening out the vortices pinned by embedded disorder in Fe-based superconductors. The approach provides an unbiased way to reveal intrinsic vortex-core states and may address puzzles on Majorana zero modes.

*This manuscript has been authored by UT-Battelle, LLC under Contract No. DE-AC05-00OR22725 with the U.S. Department of Energy. The United States Government retains and the publisher, by accepting the article for publication, acknowledges that the United States Government retains a non-exclusive, paid-up, irrevocable, world-wide license to publish or reproduce the published form of this manuscript, or allow others to do so, for United States Government purposes. The Department of Energy will provide public access to these results of federally sponsored research in accordance with the DOE Public Access Plan (http://energy.gov/downloads/doe-public-access-plan).*

In type-II superconductors, applying magnetic fields can locally break Cooper pairs and introduce magnetic vortices [1]. Understanding the electronic states at the vortex cores is crucial for determining the pairing mechanism [2]. Recent discoveries have revealed that the vortex-core state in some Fe-based superconductors can display the feature of non-Abelian Majorana zero mode (MZM), characterized by a zero-conductance peak that is robust against magnetic fields [3-8]. Vortex-core states can be studied using scanning tunneling spectroscopy (STS), where the differential conductance is proportional to the local density of states (LDOS). However, the interpretation of STS measurements is complicated by the extrinsic factors that are invisible in topography. These factors include embedded impurities, chemical inhomogeneities, or crystal disorders below the surface, which can modify the vortex-core states. [9-13]. These same embedded impurities and disorders act as pinning centers that trap vortex motion under increasing magnetic fields and distort the vortex lattice (known as bulk pinning) in a Fe-based superconductor [14-16]. Therefore, to screen out these extrinsic factors from the vortex-core states, it is necessary to correlate vortex-core states with their pinning behaviors. This forms the basis of our method for revealing the intrinsic vortex-core states.

To obtain intrinsic vortex-core states, it would be ideal to have a sample that is almost free of crystal disorders and exhibits periodic vortex lattice [17,18] due to pure inter-vortex repulsion. Unfortunately, in many samples of Fe-based superconductors, embedded chemical inhomogeneities and disorders are present and lead to distorted vortex lattices through a strong pinning mechanism [16, 19, 20]. This makes it challenging to determine which vortices display intrinsic vortex-core states. Additionally, the limited efficiency of STS experiments means that only a limited number of vortices can be examined at high spatial and energy resolution, making it difficult to decide the origins of various spectral features. In particular, a zero-bias peak induced by disorder can mimic aspects of MZM and lead to biased interpretations of STM data, which should alarm the whole community [10]. Therefore, unbiased approaches for identifying defect-free vortices with intrinsic electronic states must be developed for Fe-based superconductors.

This work combines machine learning with large STS data acquired at compromised resolution for efficiency to identify vortices without subsurface defects or disorder. The identification is based on a discovered correlation between the vortex positions and vortex-core states in Ba(Fe$_{0.96}$Ni$_{0.04}$)$_2$As$_2$, where the strong vortex pinning mechanism prevails [21]. The same vortex pinning mechanism also dominates in FeTe$_x$Se$_{1-x}$ [22], where the presence of MZM modes in the vortex-core states has been questioned [10]. Therefore, we expect this method might also be useful in related materials.

Here we start by presenting some material characterizations, including the surface structure, maps of superconducting gaps and surface defects, vortex images, and spatially averaged vortex-core states (Figs. 1-3). We then show the uncovered correlation between vortex positions and vortex-core states up to a magnetic field of 4 T (Fig. 4), forming the basis for the method of screening vortices with strong pinning centers. Finally, the overall methodology is summarized (Fig. 5).

Topographic images of Ba(Fe$_{0.96}$Ni$_{0.04}$)$_2$As$_2$ were obtained in a constant current mode by performing scanning tunneling microscopy (STM). The STM experiment and sample growth details can be found in the supplement. A topographic image of the entire 200 x 200 nm$^2$ area was captured at 4.2 K without magnetic fields and is shown in Fig. 1(a). A 2 x 1 surface reconstruction was identified from a magnified view of a subarea, as shown in Fig. 1(b), with no twin boundaries observed in the entire area studied.

Current-imaging tunneling spectroscopy (CITS) was performed over the same area as the topography in Fig. 1(a) with a lock-in amplifier modulating the bias voltage. A grid of 128 x 128 pixels of dI/dV spectra with an energy resolution of 0.25 meV was acquired at 0 T. The superconducting gap map [Fig. 2(a)] was calculated from the coherence peaks in denoised dI/dV curves, where principal component analysis (PCA) (see supplement) was used to denoise the spectra. The gap size histogram in Fig. 2(b) shows that the superconducting gap size peaks around 4 meV, consistent with previous reports [16, 23]. The surface also exhibits electronic inhomogeneity (indicating surface defects), as evidenced by a map of spectral classes [Fig. 2(c)], which distinguish between areas with and without superconducting states. The average dI/dV curves of each class are shown in Fig. 2(d).

A grid of 128 x 128 pixels of dI/dV spectra with an energy resolution of 1 meV was acquired using CITS under increasing perpendicular magnetic fields from 2 T to 4 T and 6 T, without demagnetization in between. The spectra were acquired over the same area as the topography.

Vortex images were constructed by subtracting the zero-bias map at 0 T from the maps at 2 T, 4 T, and 6 T, as shown in Fig. 3(a), (b), and (c), respectively. The zero-bias maps in nonzero fields were carefully aligned with the one at 0 T by maximizing the cross correlation of the simultaneously acquired topographic images, resulting in correlation coefficients of 0.91, 0.86, and 0.88, respectively.

Vortex locations were found by applying a peak-finding algorithm to the Gaussian-smoothed vortex images. Some vortices were found to remain in the same locations, within a 3-pixel-distance error to account for small alignment errors, as the magnetic field was increased from 2 T to 4 T and from 4 T to 6 T. Specifically, 14 out of 71 vortices in 2 T remained in the same places as the field was increased to 4 T [yellow circles in Figs. 3(a) and (b)], and 5 vortices remained in the same locations as the field was further increased to 6 T [red circles in Figs. 3(a-c)]. These unmoving vortices are likely pinned and will be further discussed in the next section.

The spatial evolution of dI/dV spectra, averaged over all the vortices, was calculated for each magnetic field [Figs. 3(d-f)]. No peak was observed at or near the Fermi level, similar to the results in Co-doped BaFe$_2$As$_2$ [16]. The causes of this absence are a topic of ongoing debate [24-27], but they will not be addressed here.

An interesting observation was that the average vortex-core states looked different as the magnetic field was increased from 2 T to 4 T and 6 T. Specifically, the dI/dV curve at the vortex centers in 2 T [the purple curve in Fig 3(d)] appeared much shallower than those in 4 T and 6 T [the purple curves in Figs 3(e) and (f)]. To exclude the possibility that these differences were caused by surface inhomogeneity, the 0 T dI/dV curves were averaged within the same circles as the vortex cores in each vortex image [Figs. 3(g-i)], revealing that they were almost identical to each other. Furthermore, no correlation between vortex positions and surface defects was found from the surface topographic image (as shown in supplementary information Fig. S1). This confirms that the difference in the average vortex-core states between 2 T and 4 T (and between 2 T and 6 T) is not due to surface inhomogeneity.

We performed a principal component analysis (PCA) on all the vortex-core states at 4 T, regardless of their distances from the vortex centers. The transformed spectral data was represented as a linear combination of the first 3 principal components (PCs) shown in Figs. 4(a-c). These first 3 PCs accounted for 91% of the variation in the entire spectral data, suggesting that all the vortices could be classified based on the spatial maps of PC weights/coefficients [Figs. 4(d-f)].

In Figs. 4(d-f), the unmoving vortices are circled in yellow as in Fig 3(b). The weights in Figs. 4(d) and (f) show random fluctuations, while in Fig. 4(e), the weights for PC2 exhibit negative values in nearly all the circled vortices and more fluctuation or positive values in the rest of the vortices. This indicates that the electronic states in the unmoving vortices are unique. The negative values in PC2 weights imply an enhanced and less concave-up dI/dV near the Fermi level, which can be confirmed by the comparison of averaged dI/dV for the circled and uncircled vortices plotted in Figs. 4(h) and (i). The sum of PC2 weights for each vortex in Fig. 4(e) shows that all 14 unmoving vortices have more negative values than the rest (Fig. 4(g)).

To investigate the influence of surface disorders, we also applied PCA to the 0 T spectra in the same regions as the vortex cores at 4 T (Fig. S2). No correlation was found between the PC weights and the circled regions, suggesting that the pinning centers are mostly independent of surface inhomogeneity or disorders, consistent with previous findings in Co-doped BaFe2As2 [16]. Instead, the reason for the distinctive vortex-core states in the unmoving vortices in 4 T field is probably due to bulk pinning [15]. Unlike highly anisotropic superconductors such as $Bi_2Sr_2CaCu_2O_{8+\delta}$ [28], surface pinning is not dominant in this case.

The PCA analysis was also applied to the 2 T and 6 T spectral data (Fig. S3). The 2 T vortex-core states showed only fluctuations in the PC weights, suggesting only one type of vortex-core state. However, in the 6 T vortex-core states, a clear separation in vortex-core states was also found, similar to the 4 T case. All 5 unmoving vortices in 6 T field showed negative PC2 weights, and some uncircled vortices also showed negative PC2 weights (Fig. S3). This could be explained by the possibility that some vortices that only appear after 4 T are trapped by embedded pinning centers while others are not. The detailed explanation for these phenomena will be discussed in the discussion.

Based on the correlation between unmoving vortices and their unique vortex-core states, we developed a method for identifying these vortices that are impacted by pinning centers beneath the surface, such as subsurface disorders. The workflow for this method is outlined in Fig. 5, demonstrating how to eliminate the vortex-core states that are related to defects and disorders both on the surface and below it. The workflow essentially outlines an algorithm that can automate the process of revealing the intrinsic vortex-core states in an unbiased way.

From the correlation between vortex-core states and positions, we can determine which vortices are free from embedded pinning sites and study their intrinsic vortex-core states. This classification is achieved by examining the dynamics of vortex nucleation, which is driven by the interplay between inter-vortex repulsion, pinning attraction, and thermal fluctuations. As demonstrated in Lorentz transmission electron microscopy studies of niobium films [29, 30], vortex nucleation in strong pinning systems occurs in three stages [31]: stage 1, where vortices form at pinning centers and occupy one pinning center each (i.e., single-vortex pinning); stage 2, where the density of vortices matches the density of pinning centers (i.e., the matching field [32, 33]) and the later-formed vortices occupy defect-free interstitial sites; and stage 3, where single-vortex pinning transitions to collective pinning and vortices move in single lines.

Since Ba(Fe$_{0.96}$Ni$_{0.04}$)$_2$As$_2$ also contains strong pinning centers [21, 34], the observed correlations can be understood in terms of these stages. The 2 T vortex lattice corresponds to stage 1, as there is only one type of vortex-core state associated with the pinned vortices. The 4 T vortex lattice is consistent with stage 2 or between stages 2 and 3, as there are two types of vortex-core states associated with pinned and interstitial vortices. The 6 T vortex lattice is mostly in stage 3, as there are fewer singly pinned vortices and a vortex image of lines of vortices is observed.

Basically, our method (Fig 5) searches for interstitial vortices in stage 2, where the field is slightly above the matching field. These vortices are away from pinning sites in the bulk and carry fewer extrinsic features.

Our approach has the potential to shed light on puzzles in other Fe-based superconductors, such as FeTe$_x$Se$_{1-x}$, which also exhibit strong pinning behaviors [22, 35]. Recently, Majorana zero modes (MZMs) were claimed to have been observed in the vortices of FeTe$_x$Se$_{1-x}$, with a persistent zero-energy mode found in a fraction of vortices under magnetic fields ranging from 0.15 to 6 T [5]. Our method, which can identify the defect-free vortices in strong pinning systems, may provide a valuable tool in understanding the intrinsic vortex-core states of this material.

In conclusion, our results demonstrate that even basic machine learning algorithms can uncover significant distinctions and correlations in STS data, offering new perspectives into the microscopic origins of vortex-core states. While supervised machine learning could also be utilized for this purpose, the choice of an appropriate model would be crucial. We propose the use of unsupervised machine learning for the comprehensive analysis of large STS data to minimize subjectivity and operator dependence in interpreting vortex-core states.

**Acknowledgements**

This work was supported by the U.S. Department of Energy (DOE), Office of Science, Basic Energy Sciences (BES), Division of Materials Sciences and Engineering and the STM experiment was conducted at the Center for Nanophase Materials Sciences (CNMS), which is a US Department of Energy, Office of Science User Facility at Oak Ridge National Laboratory.

**Author contributions**

Y.G., S.V.K. and Z.G. conceived the ideas about the data analysis. Y.G. and S.V.K carried out the machine learning. Y.G., H.M., A.R.L wrote the manuscript. Q.Z., Z.G. and M.F. conducted the STM experiments. A.S.S. grew the crystal. Y.G., H. M. and Q.Z contributed equally.

**Data availability**

The data in this study is available at https://github.com/DrYGuo/STS.

**Code availability**

The code that accompanies this study is available at https://github.com/DrYGuo/STS.

**Figures**

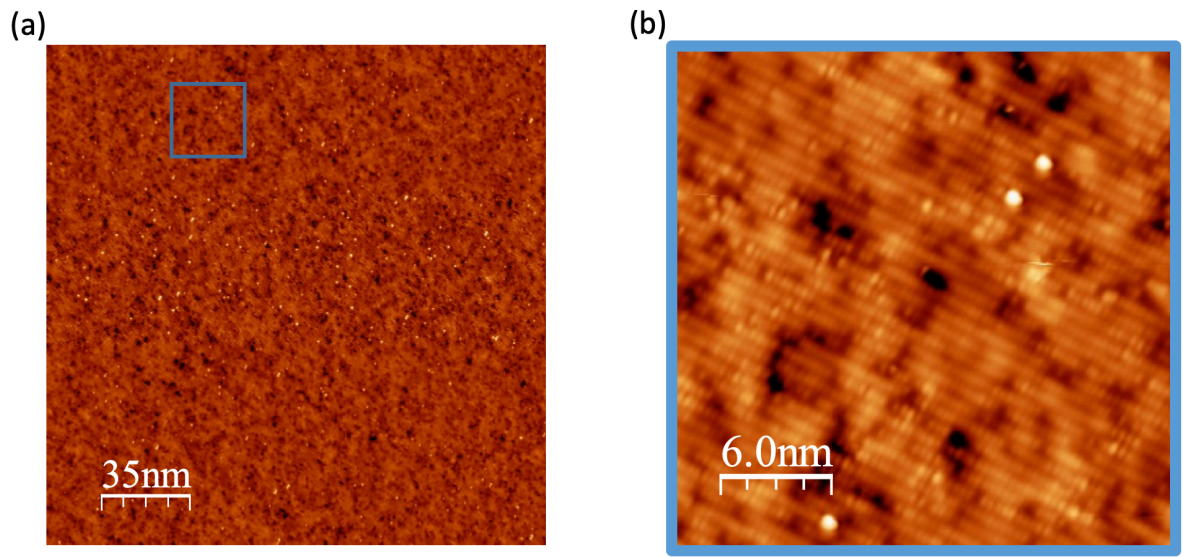

FIG. 1. Topographic images of Ba(Fe$_{0.96}$Ni$_{0.04}$)$_2$As$_2$. (a) Topographic image of the entire sample surface in this study, with the CITS measurements conducted over the same area. No twinning boundaries are visible, ruling out twin boundaries as the dominant pinning sites [40]. A subarea (indicated by the blue square) is shown in a magnified view. (b) A 2 x 1 type surface reconstruction is evident in the magnified area.

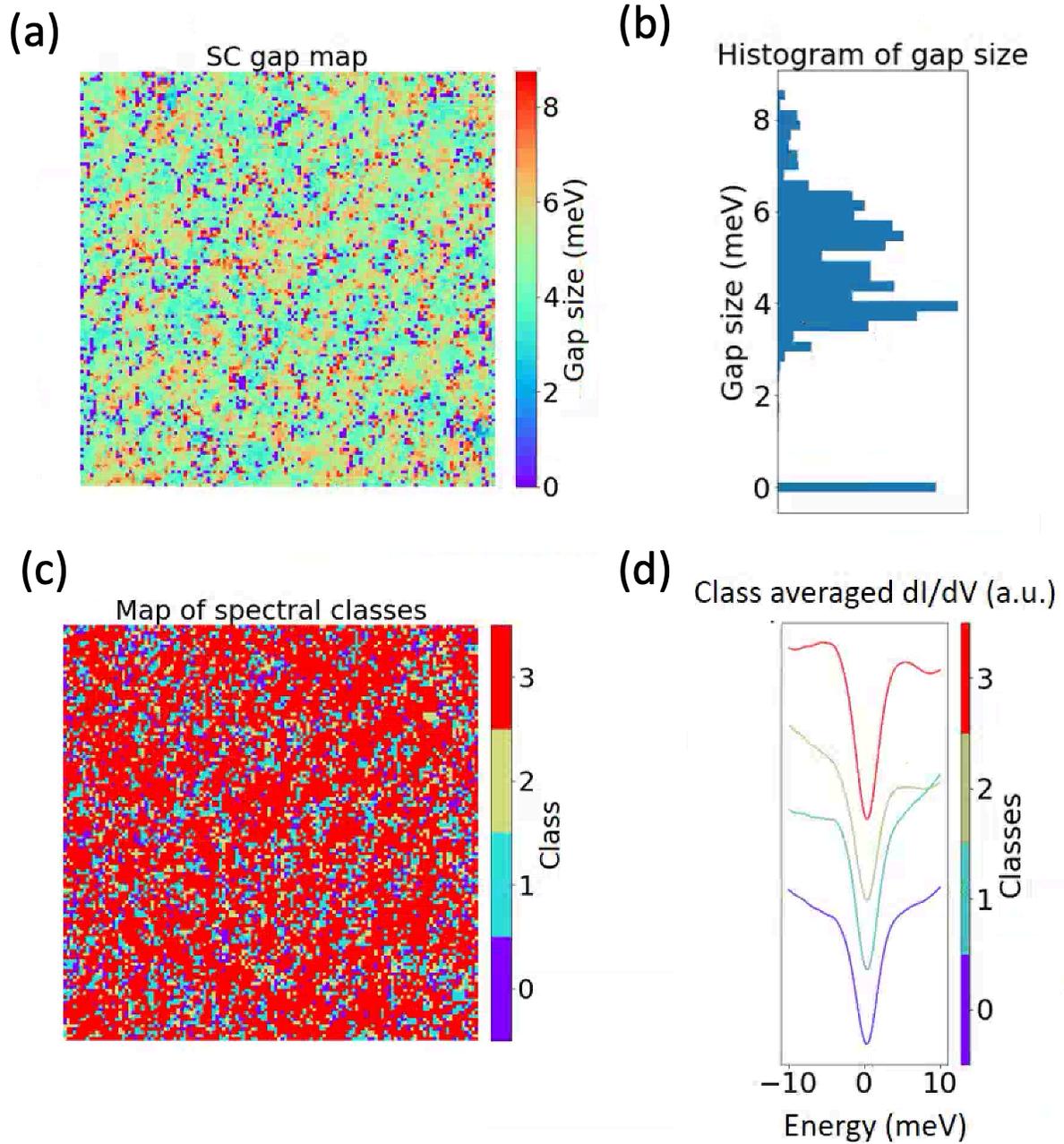

FIG. 2: Mapping of superconducting gaps and surface defect states from CITS data at 0 T. (a) A superconducting gap map of the same area as Fig 1(a). The gap size is determined as the half distance between two coherence peaks. For spectra with only one sharp coherence peak, the gap size is determined as the distance from the single coherence peak to the zero bias. (b) A histogram of the gap sizes, where spectra without coherence peaks are counted as zero. (c) A map of spectral classes, with red areas (Class 3) representing areas with complete superconducting features, as shown in the red curve in (d). This map covers the same area as Fig. 1(a). (d) Class-averaged dI/dV curves, where Class 0 has no coherence peaks or superconducting gap, Class 1 has a single coherence peak on the occupied side, Class 2 has a single coherence peak on the unoccupied side, and Class 3 has coherence peaks on both sides. In (c), 64% of the total area is occupied by spectra in Class 3.

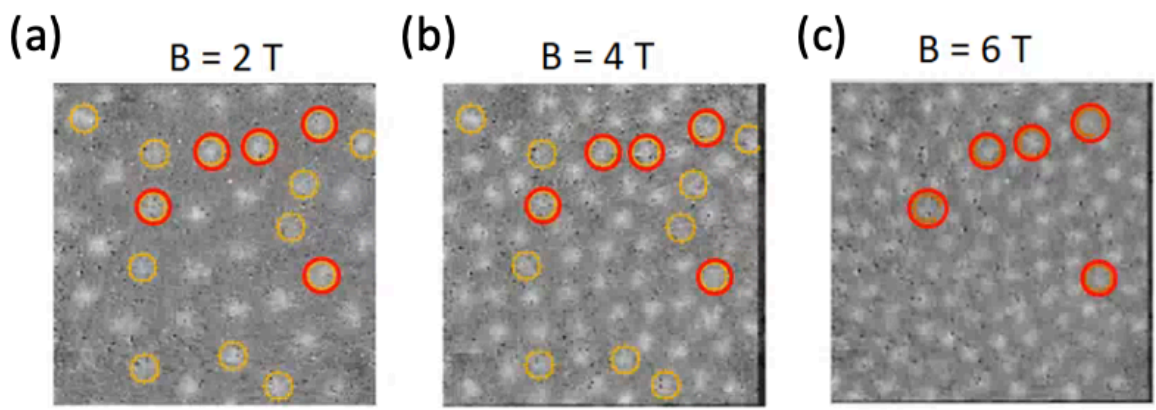
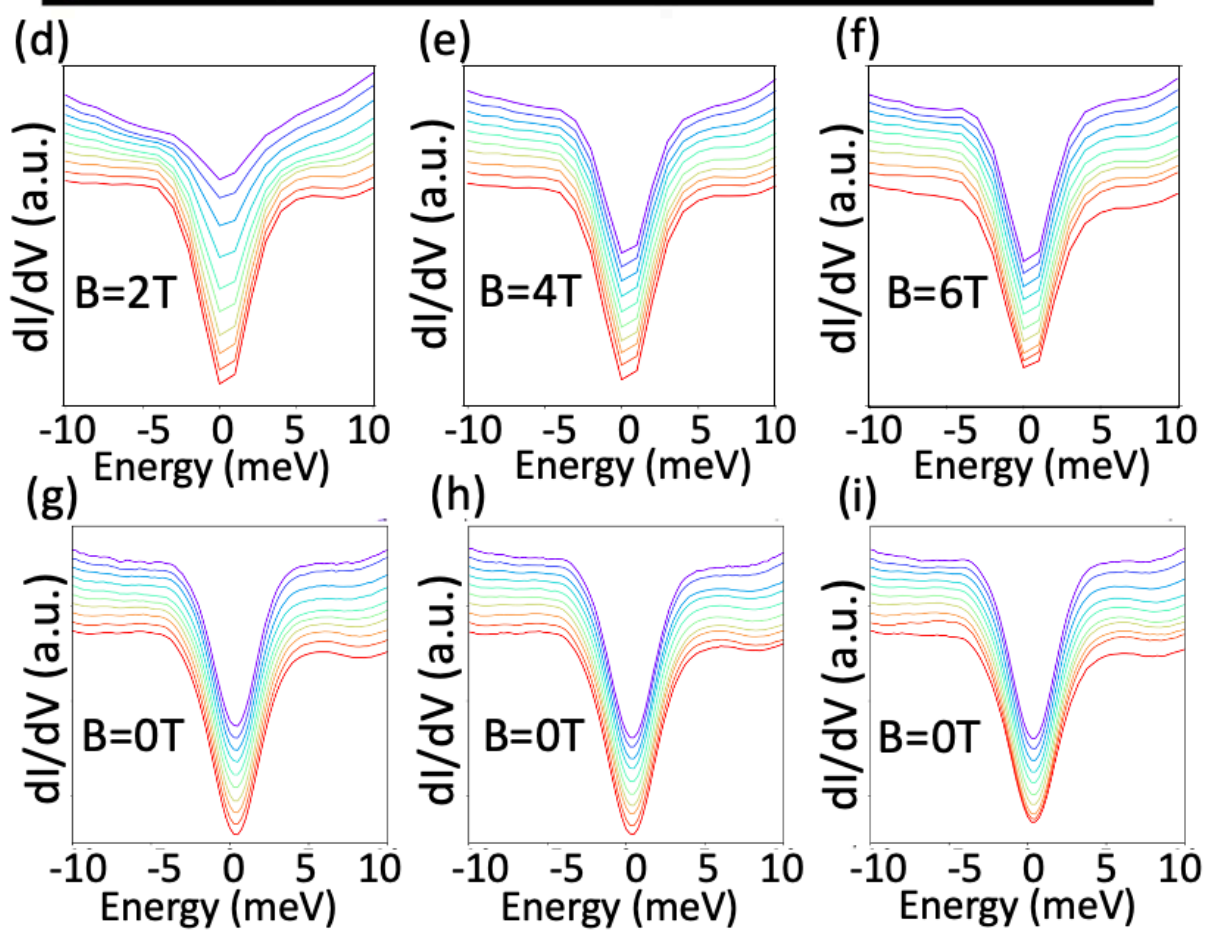
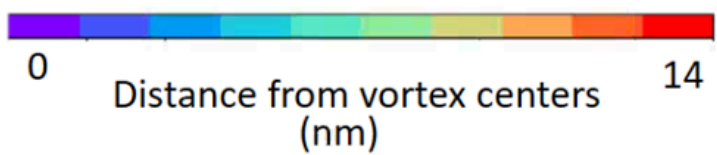

FIG. 3: Vortex images and averaged vortex cores states. (a-c) Vortex images of the same area as depicted in Fig 1(a) under magnetic fields of 2, 4, and 6 T, respectively. As the field is increased from 2 T to 4 T and further to 6 T, some vortices (circled) remain unmoving (within a radius of 3 pixel / 4.68 nm ). The vortex images are obtained by subtracting the zero-bias maps from the finite fields from the zero-field map. The zero-bias maps in non-zero fields were carefully aligned with the zero-field map by maximizing the cross-correlation of simultaneously acquired topographic images. Residual artifacts from the subtraction are visible on the margins of the vortex images. (d-f) The spatial evolution of dI/dV curves averaged over all vortices in the vortex images (a-c), respectively, reveals a significant difference at the vortex cetenters (purpule lines) between 2 T and 4 or between 2 and 6 T. (g-i) The averaged dI/dV curves at 0 T are obtained from the same locations as in (d-f), respectively. No significant difference is present in the same locations, indicating that the differences in (d-f) are not caused by surface inhomogenity.

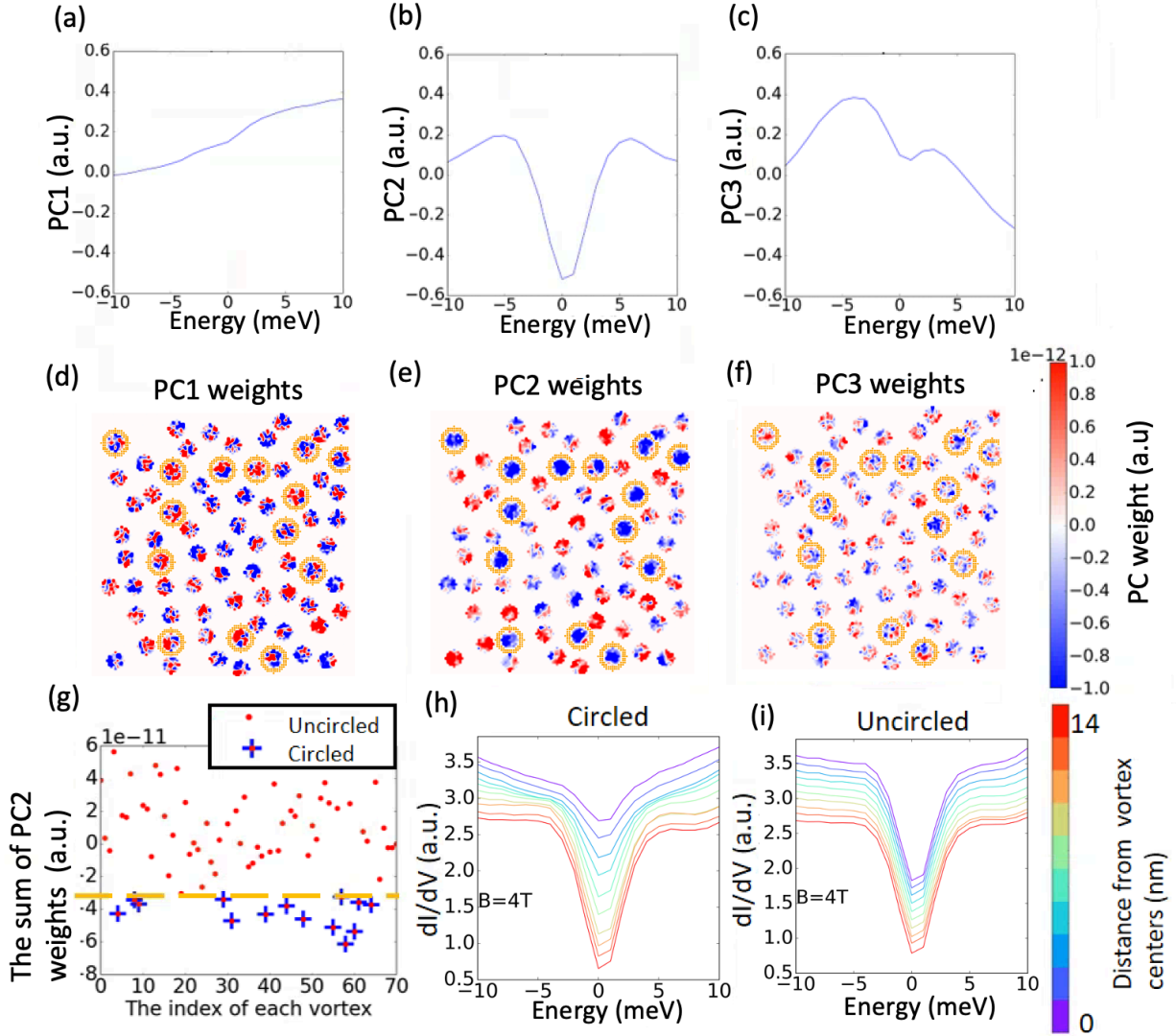

FIG. 4: A correlation between vortex positions and vortex-core states is found when applying principal component analysis (PCA) to the scanning tunneling spectroscopy (STS) data within vortices at 4 T. The vortices region is defined as a circular areas with a radius of 4 pixels (6.24 nm), slightly larger than the coherence length $\xi$ (about 4-5 nm [23]). (a-c) The first 3 principal components (PCs) of the dI/dV spectra within the vortex regions account for 91.3% of the variance in all the vortex-core states. (d-f) The maps of coefficients/weights for each PC correspond to the same area as depicted in Fig 3(b). The vortex regions are colored, with unmoving vortices circled in orange [same as in Fig 3(b)]. The PC2 weights in (e) clearly correlates with the positions of the circles. (g) The sum of PC2 weights in each vortex is plotted for 71 vortices in the 4 T field, which demonstrate a clear distinction in vortex-core states

between unmoving (circled) and moving (uncircled) vortices, due to pinning, as highlighted by the dashed yellow line. This plot highlights the statistical significance of the correlation between vortex positions and vortex-core states — the unmoving vortices all have more negative values in their summed PC2 weights than any of the rest. (h) The averaged differential conductance is shown for the circled (unmoving) vortices and (i) the uncircled (moving) vortices.

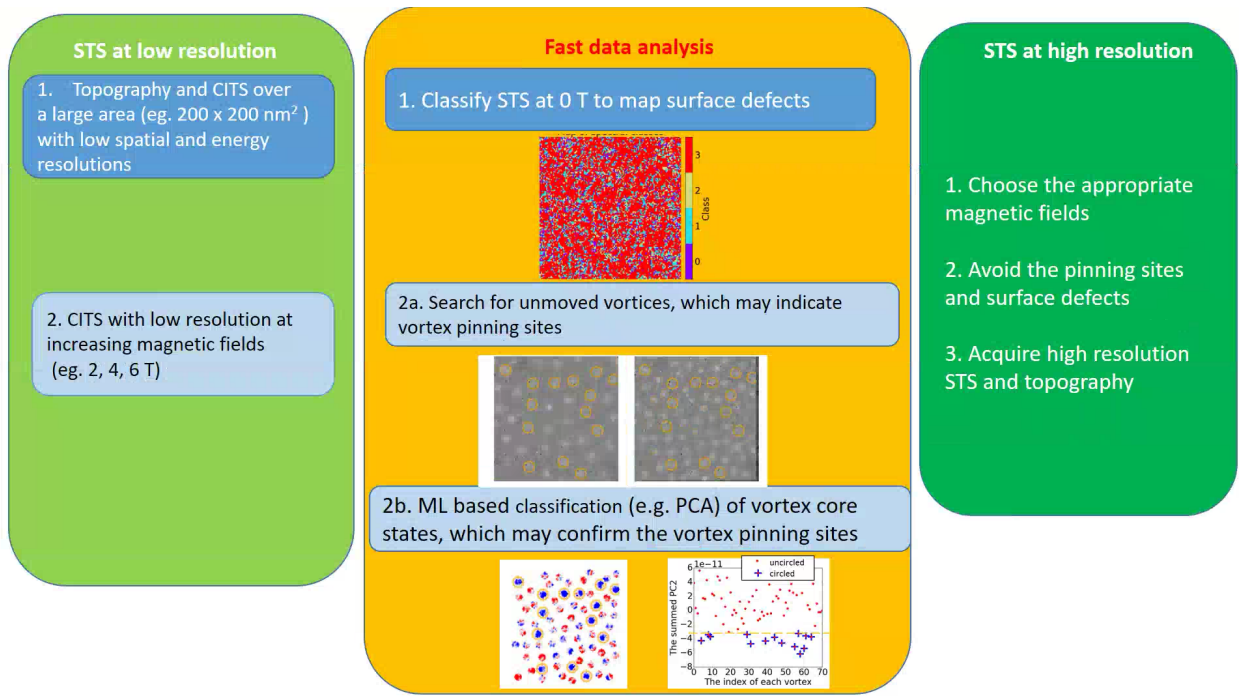

FIG. 5: A flowchart of the proposed method for identifying intrinsic vortices. The process begins with low-resolution STS data (left, light green). This data is then screened for pinned vortices and surface defects using a fast data analysis process (middle, orange). The remaining vortices are then investigated using high-resolution STS (right, dark green).

**Supplementary information:**

**Revealing intrinsic vortex-core states in the Fe-based superconductors through machine-learning-driven discovery**

Yueming Guo[1]*†, Hu Miao[2]†, Qiang Zou[1,3]†, Mingming Fu[1,4], Athena S. Sefat[2], Andrew R. Lupini[1], Sergei V. Kalinin[1]*, Zheng Gai[1]*

[1]Center for Nanophase Materals Sciences, Oak Ridge National Laboratory, Oak Ridge, Tennessee 37830, USA

[2]Materials Science and Technology Division, Oak Ridge National Laboratory, Oak Ridge, Tennessee 37830, USA

[3]Department of Physics and Astronomy, West Virginia University, Morgantown, West Virginia 26506, USA

[4]Department of Physics, OSED, Fujian Provincial Key Laboratory of Semiconductor Materials and Applications, Xiamen University, Xiamen 361005, China

**Sample growth, transport measurement and chemical analysis**

Single crystals of Ba(Fe$_{0.96}$Ni$_{0.04}$)$_2$As$_2$ were grown with the self-flux method [36, 37]. The superconducting transition temperature in 0 T magnetic field was determined to be T$_c$ = 18 K with a transition width of ΔTc < 0.5 K. The narrow transition width demonstrates the crystal quality and the cleanliness of the crystal should be comparable to previous reports. Phase purity and crystallinity were checked using Powder X-ray diffraction with an X'Pert PRO MPD diffractometer (Cu K$_{\alpha 1}$ radiation, λ=1.540598 Å). The chemical composition (Ni : Fe = 0.04 : 0.96) was determined by using a Hitachi S3400 scanning electron microscope (SEM) equipped with energy dispersive spectroscopy (EDS) at 20 kV.

**Scanning tunneling microscopy and spectroscopy**

The crystal of Ba(Fe$_{0.96}$Ni$_{0.04}$)$_2$As$_2$ was cleaved at 78 K in ultra-high vacuum (UHV) and was immediately transferred in-situ to the STM/S head which was precooled to 4.2 K. The STM/S experiments were carried out at 4.2 K using a UHV low-temperature and high field STM with a base pressure lower than 2×10$^{-10}$ Torr [22, 38]. Pt-Ir tips were prepared by mechanical cutting and were conditioned on a clean Au (111) surface. Topography, surface state and work function of Au (111) surface were checked before measurements. The STM/S was carried out in a SPECS Nanonis control system. Topographic images were acquired in a constant current mode and current-imaging tunneling spectroscopy (CITS) was performed using a lock-in amplifier to modulate the bias voltage by $dV$ of 0.1 to 1 mV at 973 Hz.

**Principal component analysis**

Principal component analysis (PCA) can reduce the dimensionality of datasets and increase the interpretability with minimal information loss. In PCA, the covariance matrix of data is decomposed into a linear superposition of eigenvectors (called principal components or PCs) and the corresponding eigenvalues (which represent the fraction of explained variance) are in descending order. Mathematically speaking,

$$x_i = \sum_{j=1}^{d} z_{ij} w_j \quad (1),$$

$$\mathbf{X}\mathbf{X}^\mathrm{T} w_j = \lambda_j w_j \quad (2),$$

where $x_i$ is the i$^{th}$ vector (in our case, a vector refers to a spectrum of dI/dV with the mean subtracted), $w_j$ is the j$^{th}$ principal component, $\mathbf{X} = \{x_i\}$, and $\lambda_i$ is the fraction of explained variance.

We applied PCA to the vortex core states by using the relevant module in scikit-learn [39]. For the 4 T vortex cores states (the original dI/dV inside the vortex regions at 4T), we obtained $\lambda_1 = 0.66$, $\lambda_2 = 0.165$, and $\lambda_3 = 0.088$ ($\lambda_1 + \lambda_2 + \lambda_3 = 91\%$), which suggests that 91% of the variation in the vortex cores states can be explained by the first 3 PCs. The 3 PCs, $w_i$, were calculated and shown in Figs 4a-c, and the PC weights, $z_{ij}$, for each individual spectrum were computed and shown in Figs 4d-f.

In generating the gap map in Fig 2a, we also applied PCA to smooth the dI/dV at 0 T before locating the coherence peaks.

**Supplementary Figures**

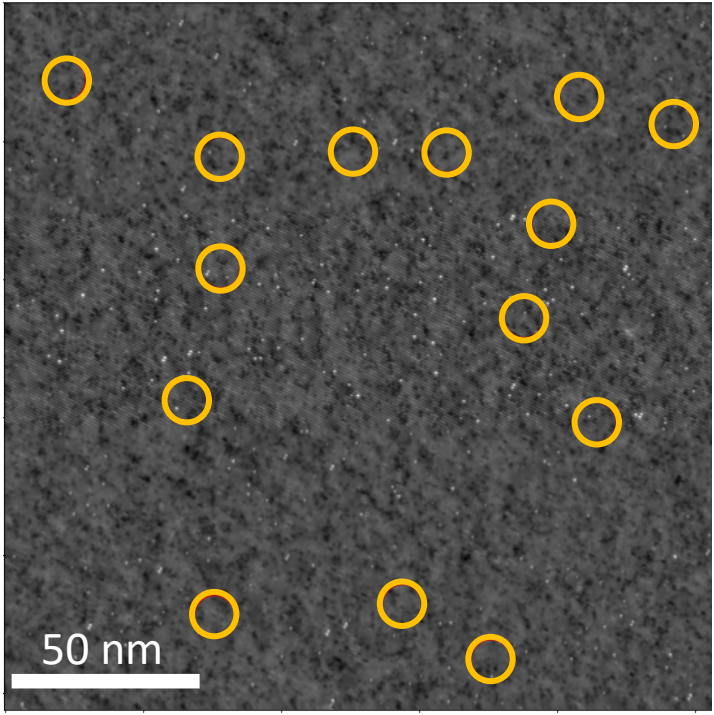

Fig S1. The locations of the 14 unmoving vortices are labelled in the topographic image. No correlation between the vortex positions and surface defects was found.

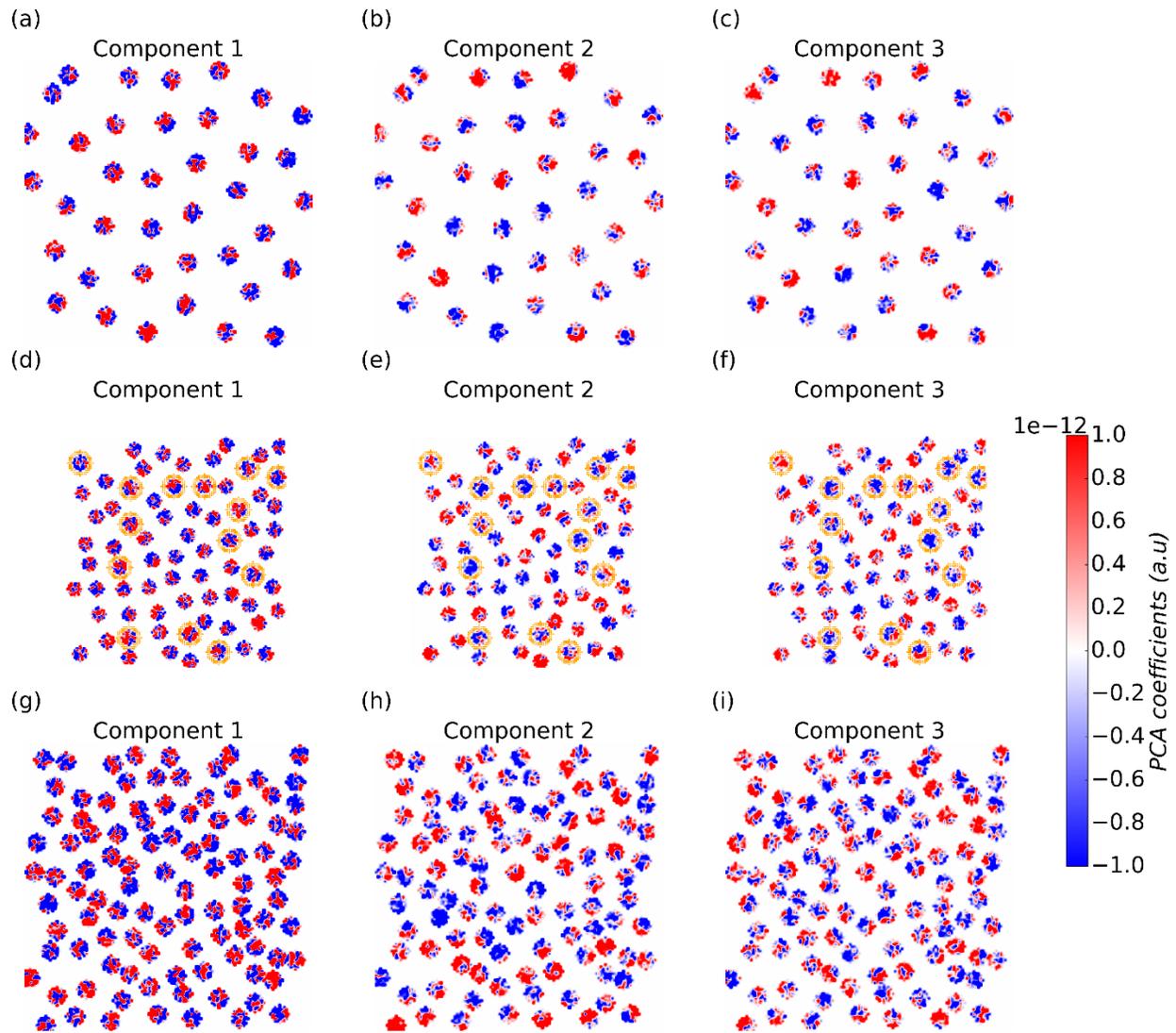

Fig S2. The principal component analysis (PCA) of the dI/dV spectra at B = 0 T. The 0 T dI/dV data from different locations were input to the PCA: from regions inside the vortices at (a-c) 2 T, (d-f) 4 T, and (g-i) 6 T. The weights of the first three principal components were plotted. In (e), there is no clear correlation between the circles (which labels the unmoving vortices) and the PC2 weights. This is in contrast to the correlation observed in Fig. 4e.

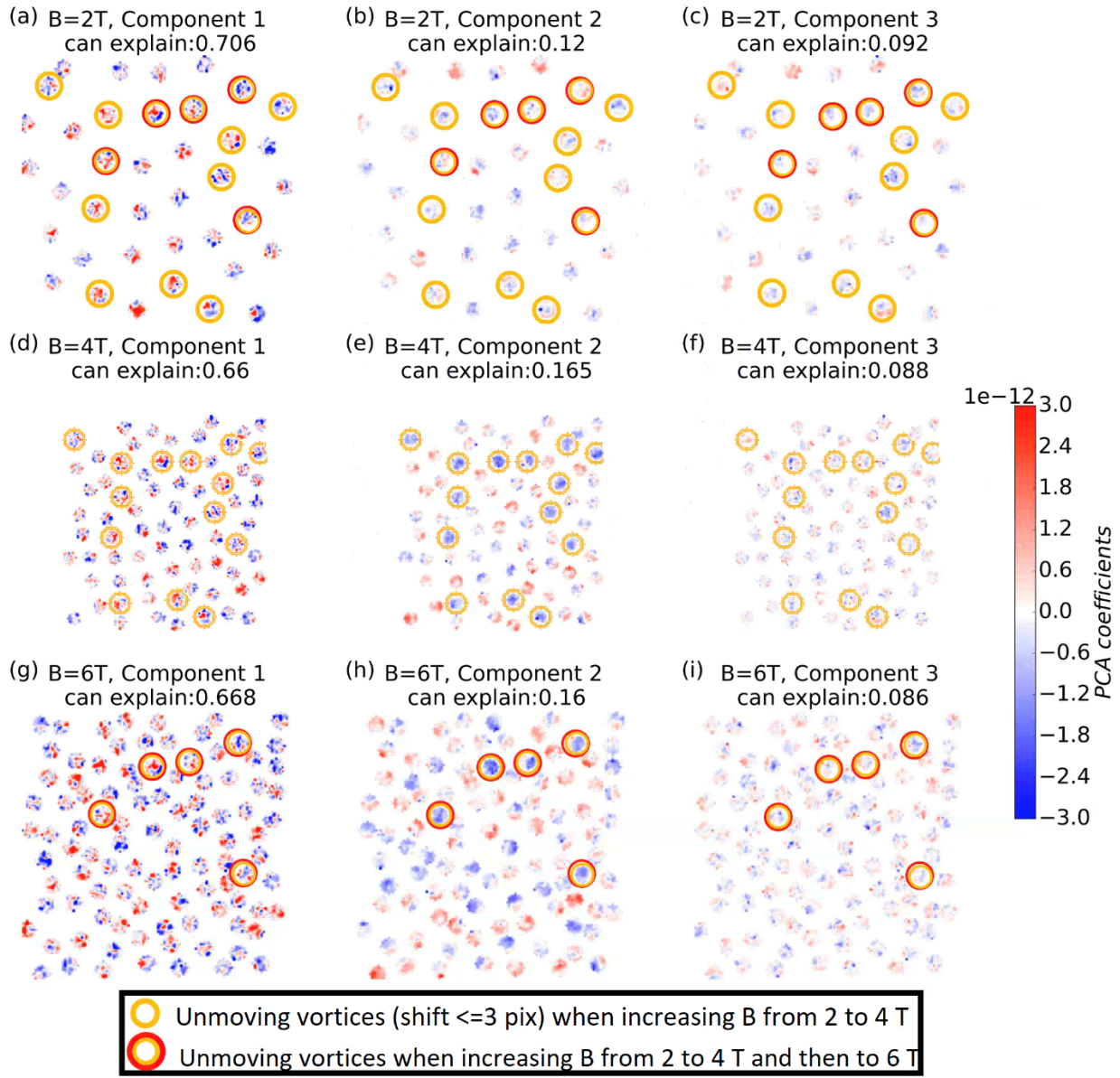

Fig S3. The principal component analysis of all the vortex core states in different magnetic fields: (a-c) 2 T, (d-f) 4 T, and (g-i) 6 T. The principal component weights for all the vortex cores states are presented. The principal components (the eigenvector basis for the curves) are similar to those in Figs. 4a-c.